\documentclass[twocolumn, times]{aastex631}

\usepackage{amsmath, latexsym, color, verbatim}

\shorttitle{3D Drizzle for JWST}
\shortauthors{Law et al. 2023}

\newcommand{\ra}{\textrm{RA}}
\newcommand{\dec}{\textrm{DEC}}

\begin{document}

\title{A 3D Drizzle Algorithm for JWST and Practical Application to the MIRI Medium Resolution Spectrometer}

\author[0000-0002-9402-186X]{David R.\ Law}
\affiliation{Space Telescope Science Institute, 3700 San Martin Drive, Baltimore, MD, 21218, USA}

\author[0000-0002-9288-9235]{Jane E.\ Morrison}
\affiliation{Steward Observatory, University of Arizona, Tucson, AZ, 85721, USA}

\author[0000-0003-2820-1077]{Ioannis Argyriou}
\affiliation{Institute of Astronomy, KU Leuven, Celestijnenlaan 200D, 3001 Leuven, Belgium}

\author[0000-0001-8718-3732]{Polychronis Patapis}
\affiliation{Institute for Particle Physics and Astrophysics, ETH Zurich, Wolfgang-Pauli-Str 27, 8093 Zurich, Switzerland}

\author[0000-0002-7093-1877]{J. \'Alvarez-M\'arquez}
\affiliation{Centro de Astrobiolog\'{\i}a (CAB), CSIC-INTA, Ctra. de Ajalvir km 4, Torrej\'on de Ardoz, E-28850, Madrid, Spain}

\author[0000-0002-0690-8824]{Alvaro Labiano}
\affiliation{Telespazio UK for the European Space Agency, ESAC, Camino Bajo del Castillo s/n, 28692 Villanueva de la Ca\~nada, Spain}
\affiliation{Centro de Astrobiolog\'ia (CAB), CSIC-INTA, ESAC, Carretera de Ajalvir km4, 28850 Torrej\'on de Ardoz, Madrid, Spain.}

\author[0000-0002-1368-3109]{Bart Vandenbussche}
\affiliation{Institute of Astronomy, KU Leuven, Celestijnenlaan 200D, 3001 Leuven, Belgium}

\begin{abstract}

We describe an algorithm for application of the classic `drizzle' technique to produce
3d spectral cubes using 
data obtained from the slicer-type integral field unit (IFU) spectrometers
on board the James Webb Space Telescope.  This algorithm relies upon the computation
of overlapping volume elements (composed of two spatial dimensions and one spectral dimension) between
the 2d detector pixels and the 3d data cube voxels, and is greatly simplified
by treating the spatial and spectral overlaps separately at the cost of just 0.03\% in
spectrophotometric fidelity.
We provide a matrix-based formalism for the computation of spectral radiance, variance,
and covariance from arbitrarily dithered data and comment on the performance of this algorithm for
the Mid-Infrared
Instrument's Medium Resolution IFU Spectrometer (MIRI MRS).  We derive a series
of simplified scaling relations to account for covariance between cube spaxels in spectra
extracted from such cubes, finding multiplicative factors ranging from 1.5 to 3 depending on the
wavelength range and kind of data cubes produced.
Finally, we discuss how undersampling produces
periodic amplitude modulations in the extracted spectra in addition to those naturally produced by fringing within the instrument; reducing such undersampling artifacts below 1\%
requires a 4-point dithering strategy and spectral extraction radii of 1.5 times the PSF FWHM or greater.

\end{abstract}

\keywords{infrared spectroscopy --- Astronomy data analysis}


\section{Introduction}

The James Webb Space Telescope (JWST) contains two integral-field unit (IFU) spectrographs
operating at near-infrared and mid-infrared wavelengths.
The Near Infrared Spectrograph \citep[NIRSpec;][]{boker22} provides IFU spectroscopy throughout
a $3 \times 3$ arcsec field from $\lambda\lambda 0.6 - 5.0 \mu$m, while the Mid Infrared Instrument
Medium Resolution Spectrometer \citep[MIRI-MRS;][]{wells15,wright23,argyriou23} provides IFU spectroscopy
throughout a $3.2 \times 3.7$ - $6.6 \times 7.7$ arcsec field from $\lambda\lambda \, 4.9 - 27.9 \mu$m \citep{labiano21}.
Both IFUs are of slicer-type design, meaning that they sample the field of
view with an optical image slicer that disperses spectra on a detector in a manner
akin to multiple adjacent slit-type apertures.


A key part of the data processing pipeline for both MIRI and NIRSpec is the reformatting of calibrated two-dimensional (2d) spectral data from multiple dithered observations
into three-dimensional (3d) rectified data cubes on a regularly-sampled grid. Such data cubes are a convenience that greatly simplifies later scientific analyses as they combine arbitrarily dithered observations into a single data product and obviate the need for a given user to know about the complex distortion solution of the instrument.
However, one-dimensional (1d) spectra extracted from such cubes can also contain artifacts produced by the resampling of the spectral data, and some science programs (e.g., observations of unresolved point sources) may
achieve higher SNR, better spectral resolution, or lower covariance
by extracting 1d spectra directly from the 2d data.


Fundamentally, a cube building algorithm converts between detector {\bf pixels} (i.e, 2d elements of the detector focal plane array) and rectified data cube {\bf voxels} (i.e., 3d volume elements with two spatial
dimensions and one spectral dimension).  
As typically implemented within the FITS image data format,
such voxels have constant spatial size throughout a given data cube but can have varying spectral width (e.g., in the case of non-linear wavelength solutions).
It is likewise common to refer to individual {\bf spaxels} within such data cubes, where
a spaxel correponds to a given 2d spatial element and consists of many individual
voxels stretching away in the spectral domain.

In the present contribution, we describe our implementation of a 3d drizzle algorithm
for the specific use case of the JWST MIRI and NIRSpec IFUs.  We note that
this is not the first time that such algorithms have been developed;
\citet{smith07}, \citet{regibo12}, \citet{sharp15}, and \citet{weilbacher20} have all discussed its application
to Spitzer/IRS, Herschel/PACS, AAO/SAMI, and VLT/MUSE respectively and similar schemes using other weighting functions have been presented previously for many other ground-based IFUs \citep[see, e.g.,][and references therein]{sanchez12, law16}. However, the specifics of the implementation often differ based on the characteristic
properties of each instrument; we therefore present the JWST-specific algorithm here
along with an analysis of its implications for data quality.
While we focus primarily on the JWST MIRI MRS IFU, most of our discussion
and conclusions are relevant for the JWST NIRSpec IFU as well (see \S \ref{nrs_appendix.sec}).

We organize this manuscript as follows.
In \S \ref{ifuoverview.sec} we review the basic design characteristics
of the MIRI MRS IFU.  In \S \ref{drizzle.sec} we describe
the algorithm used by the STScI JWST data reduction pipeline, along with some
practical notes for implementation of the algorithm in a pipeline-style environment.
We assess the covariance properties of the resulting data cubes for the MIRI MRS
in \S \ref{covar.sec}, providing a set of scale factors that can be applied to the
pipeline-generated data cubes to ensure proper variance characteristics of
extracted spectra.  Finally, we discuss the impact of the severe spatial and spectral undersampling
of the MRS on the composite data cubes in \S \ref{resampling.sec}, along with
recommendations for observing and data analysis strategies to mitigate the impact
of these effects on scientific data.
We summarize our conclusions in \S \ref{summary.sec}.



\section{Overview of the MIRI IFU and Data Pipeline}
\label{ifuoverview.sec}

\begin{figure*}[!]
\epsscale{1.2}
\plotone{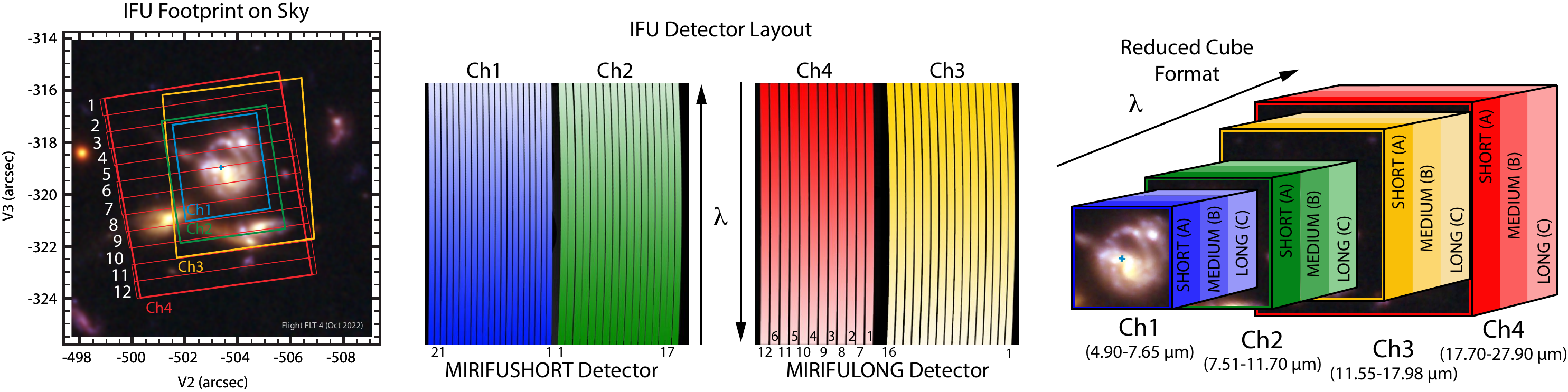}
\caption{Schematic overview of the MIRI MRS IFU.  Left-hand panel: Footprint of the IFUs for Channels 1-4 (blue, green, orange, and red regions respectively) using their A grating setting.  Individual slices have been drawn and numbered for Channel 4A.  Middle panel: Layout of the dispersed spectra on the two Si:As detectors.  Channels 1/2 and 3/4 share detectors, and the dispersion direction switches between the two.  Slices that are adjacent on the sky are interleaved in the detector plane; numbers illustrate the schematic layout.  Right panel: Composite data cube reconstructed from the dispersed spectra illustrating the wavelength coverage of each of the twelve MRS spectral bands.  Since the IFU field of view grows with wavelength this data cube is in practice akin to a stepped pyramid.
}
\label{overview.fig}
\end{figure*}

The MIRI MRS \citep{wells15} consists of four cospatial IFUs that jointly cover the wavelength range 4.9 - 27.9 $\mu$m using a series of dichroic
beam splitters and gratings to disperse light across a pair of arsenic-doped silicon (Si:As) impurity band conduction detectors with 1032 x 1024
pixels each \citep{rieke15}.  This is achieved using a pair of dichroic and grating assembly (DGA) wheels to select between
three grating settings (A/B/C\footnote{Here we use the A/B/C naming scheme; elsewhere some references instead refer to these as the SHORT/MEDIUM/LONG settings respectively.}); each observation obtains data from all four IFUs simultaneously at a single setting.
There are thus twelve sub-bands (A/B/C for each of Channels 1-4) that make up the full MRS wavelength range.  Longer wavelength
IFUs have progressively larger footprints ranging from $3.2 \times 3.7$ arcsec in Channel 1 to 
$6.6 \times 7.7$ arcsec in Channel 4.  As illustrated in Figure \ref{overview.fig}, it is the task of the IFU cube building algorithm
to create a rectified data cube (strictly more of a step-pyramid) from the dispersed spectra of each of the individual fields of view.

The on-orbit performance characteristics of the MRS have been presented by \citet[][]{argyriou23} and references therein; we summarize here those aspects
relevant to the construction and evaluation of composite data cubes.

\begin{figure}[!]
\epsscale{1.1}
\plotone{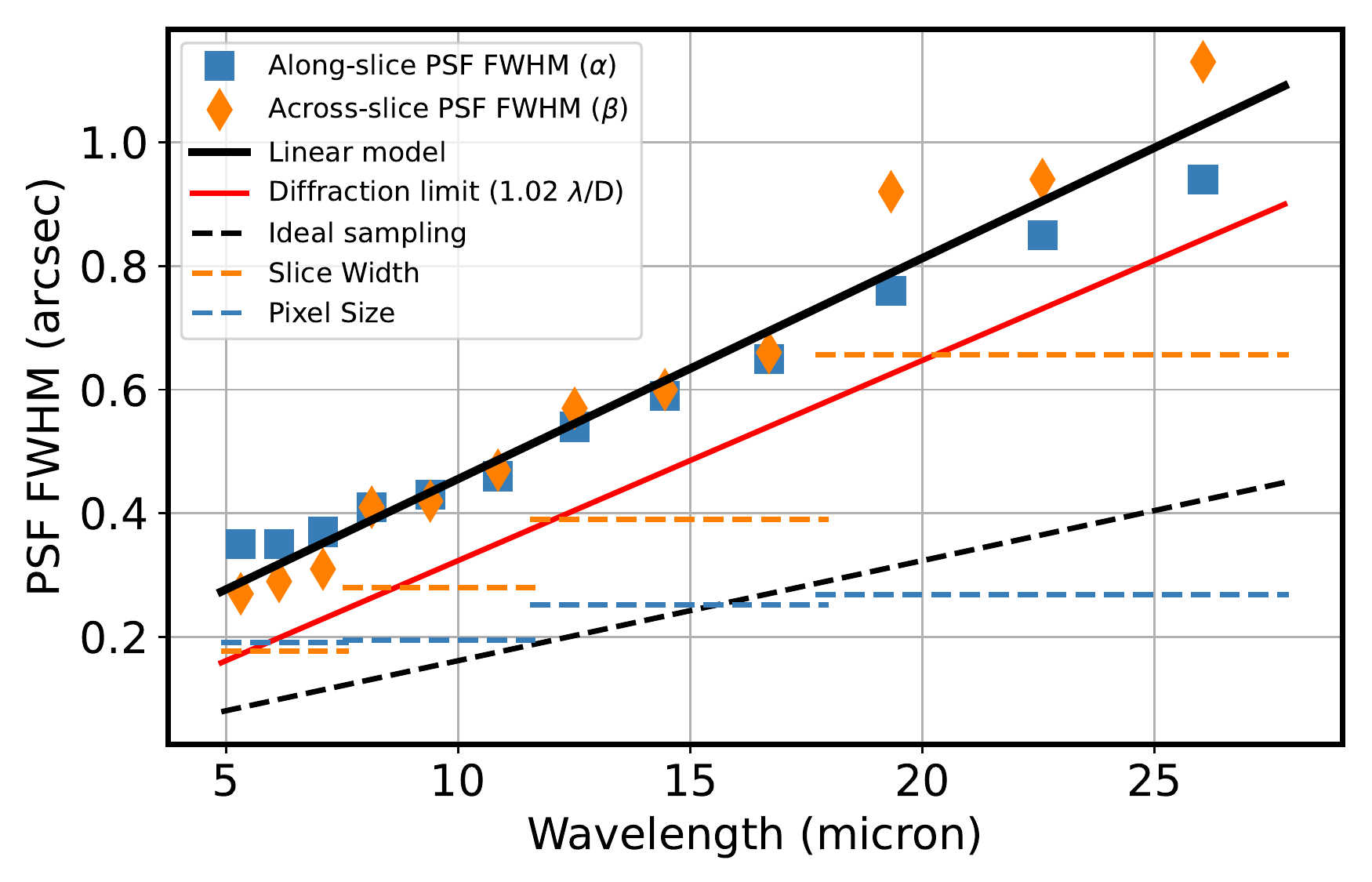}
\caption{MRS PSF FWHM in the along-slice (blue squares) and across-slice (orange diamonds) directions recovered from drizzled data cubes of point sources observed during commissioning \citep{patapis23b}. The solid black line represents a linear fit to the observed data.  The solid red line represents the theoretical diffraction limit for a 6.5m mirror, while the dashed black line represents Nyquist sampling (defined as two samples per diffraction-limited FWHM).  The dashed blue and orange lines represent the sampling provided in the along-slice direction by the detector pixel scale and in the across-slice direction by the IFU slicer width respectively.
}
\label{multiwave.fig}
\end{figure}

In order to maximize the spectral resolution and field of view
for a fixed number of detector pixels, the MRS is significantly spatially
undersampled\footnote{The MRS is, to a lesser extent, spectrally undersampled as well.}
by design and does not reach the Nyquist-sampling minimum of at least two 
samples per spatial resolution element (Figure \ref{multiwave.fig}) required to 
faithfully reconstruct the original signal.
In the IFU image slicer along-slice direction ($\alpha$) the sampling
is set by the detector pixel size, which undersamples the PSF for
$\lambda < 15 \mu$m, critically samples the PSF for $15 < \lambda < 20 \mu$m, and oversamples
the PSF for $\lambda > 20 \mu$m (dashed blue lines).  In the across-slice direction
($\beta$) the sampling is set by the width of the IFU slicer optics which undersample the PSF at all wavelengths (orange dashed lines).
These slice widths have thus been carefully chosen \citep{wells15} such that a single across-slice dither
can simultaneously achieve half-integer offsets with respect to the slicers in each
of the four MRS channels.\footnote{I.e., 0.97'' is 5.5, 3.5, 2.5, and 1.5 times the slice
width in Channels 1-4 respectively.}
Such dithering is essential in order to be able to recover spatial and spectral resolution lost due to the convolution
of the native PSF delivered by the JWST focal optics with the MRS slicer and pixel response functions.

Nonetheless, we note that the measured widths of bright point sources observed during 
JWST commissioning do not reach the theoretical diffraction limit, as illustrated by 
Figure \ref{multiwave.fig} (blue and orange points).
As discussed at length by \citet{argyriou23} and \citet{patapis23b}, this is due not just to undersampling but also to internal scattering within the MRS detectors, which broadens the effective PSF by multiple internal reflections of the wavefront.
For the purposes of the present contribution, we neglect the differences between the along-slice and across-slice PSF and
describe the average PSF FWHM of the MIRI MRS data as a linear function of wavelength $\lambda$
(solid black line in Figure \ref{multiwave.fig}) given by:
\begin{equation}
\theta = 0.033 \, (\lambda/\textrm{micron}) + 0.106 \, \textrm{arcsec}
\label{psf.eqn}
\end{equation}


\begin{figure*}[!]
\epsscale{1.1}
\plotone{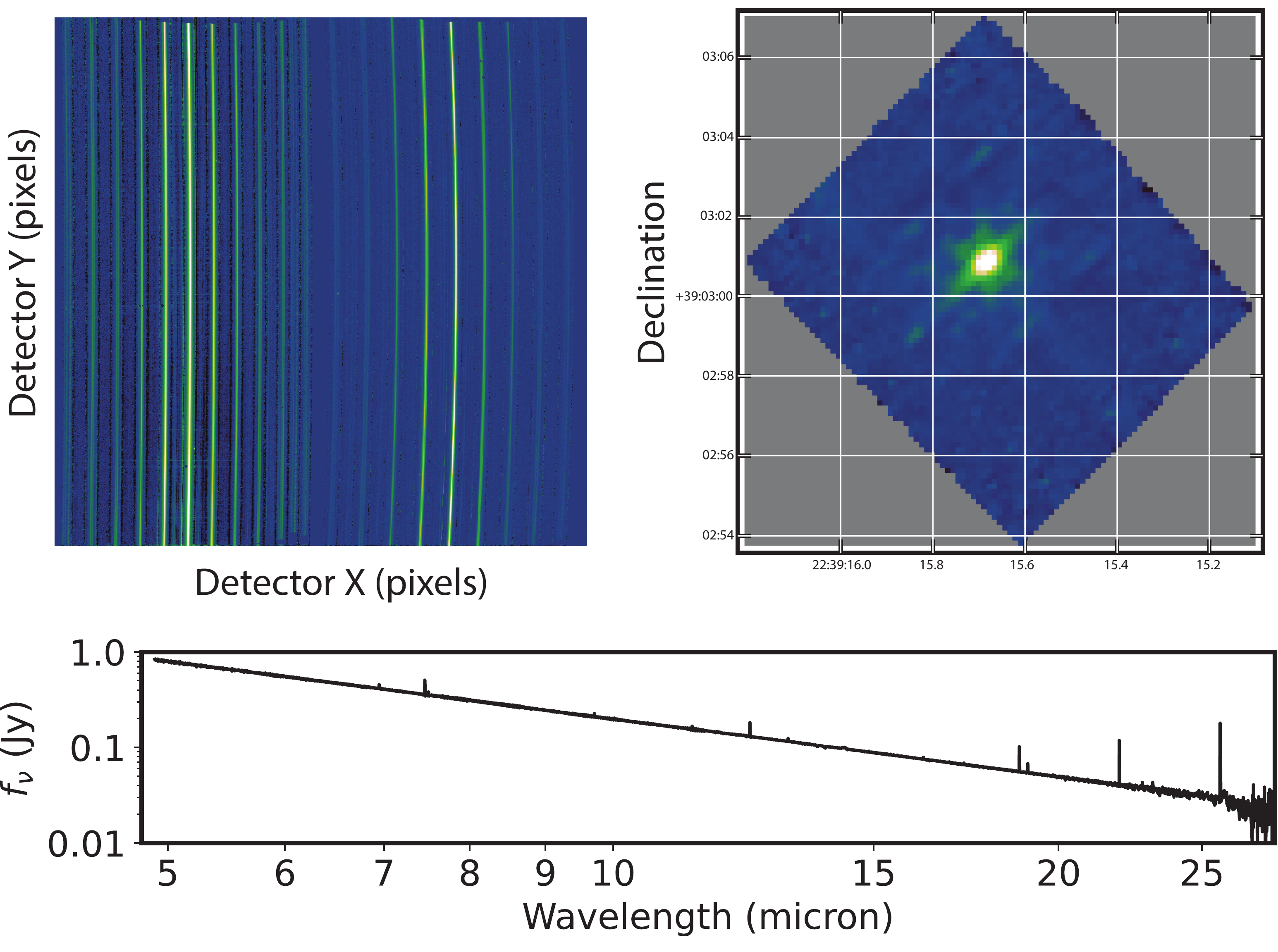}
\caption{Example MRS mosaiced observations of the bright O9V star 10 Lac (JWST Program ID 1524).
The top left panel shows the flux-calibrated 2d detector-level data for
Channels 1A+2A, the top right panel shows a 5 $\mu$m plane of the Ch1A
drizzled data cube, and the lower panel shows the 4.9 - 27.9 $\mu$m 
spectrum extracted from the drizzled data cube. The 2d detector image
and 3d cube image are both in calibrated surface brightness units and are
shown here with an identical logarithmic stretch. Values in the data cube that fall outside the IFU footprint or otherwise have no valid data (gray regions in the top-right panel) are set to NaN by default by the JWST pipeline.  Faint artifacts surrounding the source are due to imperfections in the cruciform artifact correction 
\citep[see discussion by][]{patapis23b}.
}
\label{10lac.fig}
\end{figure*}

The JWST data pipeline \citep{bushouse22} consists of three primary stages.  In the first stage {\sc calwebb\_detector1} \citep{morrison23}
the raw detector ramps are corrected for cosmic rays and a variety of electronic artifacts to produce a series of uncalibrated slope images
giving the count rate in each detector pixel in units of DN/s.  In the second stage {\sc calwebb\_spec2} these uncalibrated slopes
have the MIRI MRS wavelength calibration \citep{argyriou23} and distortion model \citep{patapis23a} attached, are corrected for the MIRI
cruciform artifact \citep{patapis23b}, flatfielded, flux calibrated, and corrected for fringes produced by
gain modulations resulting from internal reflections within the MRS detectors \citep{argyriou20,mueller23, kavanagh23}.  These calibrated slope
images are provided in units of surface brightness (MJy/sr), and have both science (SCI), uncertainty (ERR), and data quality (DQ) extensions. The task of IFU cube building is thus to take these flux-calibrated detector images and combine multiple dithered images into composite 3d science, uncertainty, and data quality cubes. 
This task occurs within the third stage of the JWST pipeline {\sc calwebb\_spec3}, along with other steps that can perform background subtraction and outlier detection based on the multiple dithered exposures.\footnote{Strictly, cube building also occurs within the
{\sc calwebb\_spec2} stage as well to create small preview cubes based on individual exposures.}
1d spectra (e.g., Figure \ref{10lac.fig}) can then easily be extracted from these cubes by applying standard 2d aperture photometry techniques to each wavelength plane, with suitable corrections for aperture losses given the reconstructed MRS PSF.

At the present time the JWST pipeline includes two method for building IFU data cubes: the 3d drizzle approach that is the subject of the present contribution and an alternative based on an exponential Modified-Shepard method (EMSM) weighting function.  While we concentrate our discussion on 3d drizzle, we compare briefly against the EMSM approach in \S \ref{emsm.sec}.


\section{The 3D Drizzle Algorithm}
\label{drizzle.sec}

\subsection{Base algorithm}

As outlined in the original case for 2d imaging by \citet{fh02}, the drizzle algorithm is based around projecting the natively-sampled detector pixels onto a common output grid and allowing the pixel
intensities to `drizzle' down onto the output grid using weights given by the
fractional area overlap between the input and output pixels. Similarly, in the 3d case we need to project the 2d
detector pixels to their corresponding 3d volume elements and allocate
their intensities to the individual voxels of the final data cube according
to their volumetric overlap.

The volumetric overlap of two arbitrarily-oriented hexahedra is a
non-trivial calculation, but can be simplified first by the assumption that
our coordinate frame is aligned with the cube voxels.  In this case, our
task reduces to computing the overlap between the irregular projected volumes
of the detector pixels and the regular grid of cube voxels which for simplicity
we assume corresponds to the world coordinates ($\ra, \dec, \lambda$).\footnote{This represents the `skyalign' orientation provided by the JWST pipeline, but cubes can also be produced with arbitrary rotation angles since the actual rectilinear grid is a cartesian plane tangent to  the spherical world coordinate system.}

\begin{figure*}[!]
\epsscale{1.1}
\plotone{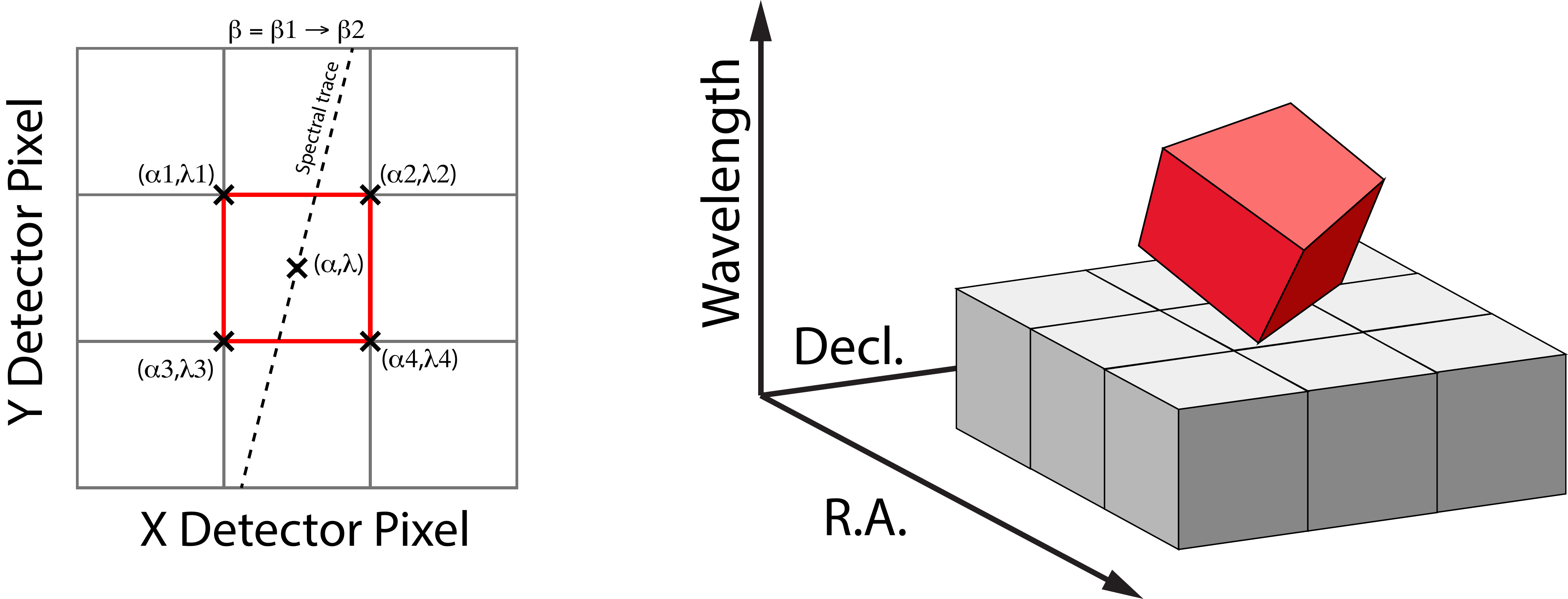}
\caption{Left: General case detector diagram in which the dispersion axis is tilted with respect to the detector columns/rows and the four corners of a given pixel (bold red outline) each have different wavelengths $\lambda$ and along-slice coordinates $\alpha$.  Right: Projection of this generalized detector pixel into the volumetric space of the final data cube.  The red hexahedron represents the detector pixel, where the three dimensions are set by the along-slice, across-slice, and wavelength coordinates.  The regular grey hexahedra represent voxels in a single wavelength plane of the data cube.
For clarity the cube voxels are shown aligned with the (R.A., Decl.) celestial coordinate frame, but this choice is arbitrary.
}
\label{diagram1.fig}
\end{figure*}

Unlike in the imaging case, the detector pixels illuminated by our slicer-type
IFUs contain a mixture
of degenerate spatial and spectral information.
The spatial extent in the along-slice direction ($\alpha$) and
the spectral extent in the dispersion direction ($\lambda$) both vary continuously within the dispersed image of a given slice
in a manner akin to a traditional slit spectrograph and are sampled by the detector pixels $(x,y)$.
In contrast, the spatial
extent in the across-slice direction ($\beta$) is set by the IFU
image slicer width and changes discretely between slices.
The four corners of a detector pixel thus define a tilted hexahedron in ($\alpha$, $\lambda$) space with the front and back faces of the polyhedron defined
by the lines of constant $\beta$ created by the IFU slicer.  ($\alpha$, $\beta$) is itself rotated (and incorporates some degree of optical distortion) 
with respect to world coordinates (\ra,\dec) and thus the volume element defined by a detector pixel is rotated in a complex manner with respect to the cube voxels
(Figure \ref{diagram1.fig}).

\begin{figure*}[!]
\epsscale{1.1}
\plotone{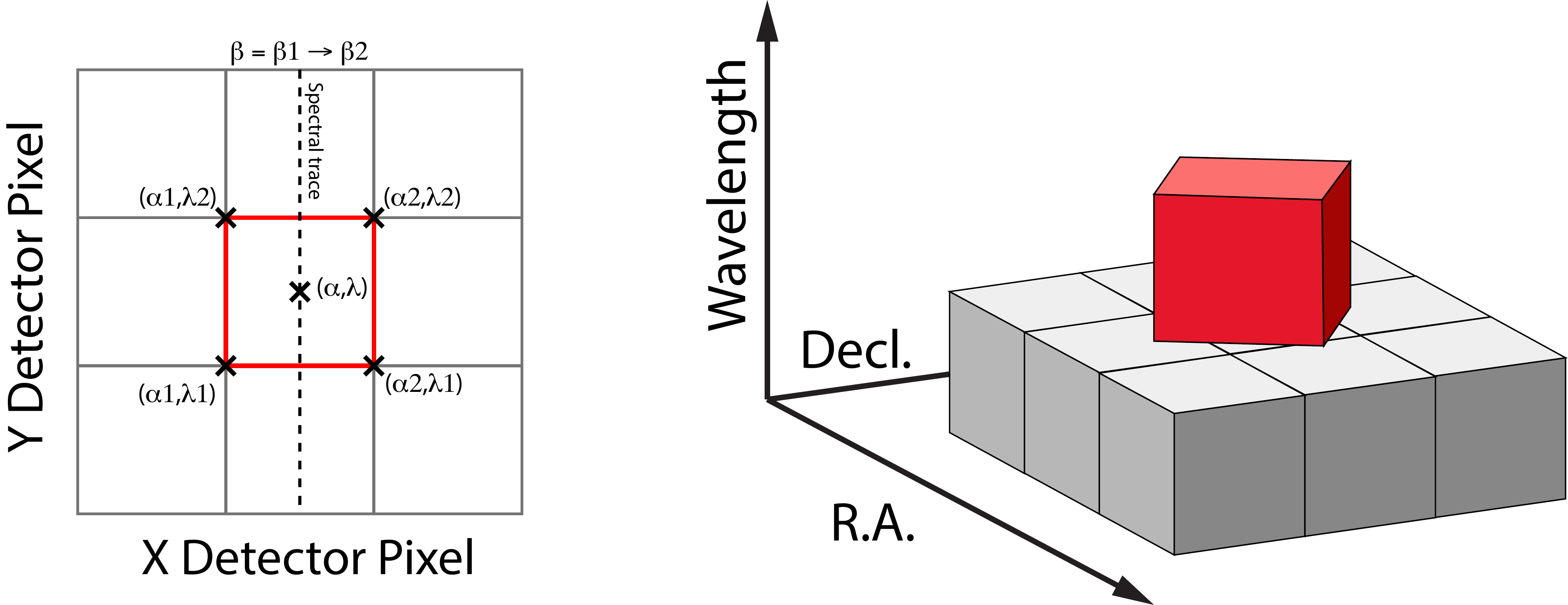}
\caption{As Figure \ref{diagram1.fig} but representing the simplified case in which 
the spectral dispersion is assumed to be aligned with detector columns and the spatial distortion constant for all wavelengths covered by a given pixel.  This assumption reduces the computation of volumetric overlap between red and 
grey hexahedra to separable 1d and 2d computations.
}
\label{diagram2.fig}
\end{figure*}

The iso-$\alpha$ and iso-$\lambda$ directions are not perfectly orthogonal to each other, and are similarly tilted with
respect to the detector pixel grid.  However,
since iso-$\alpha$ is nearly aligned with the detector Y axis (mean tilt of 1$\pm 0.8^{\circ}$ across all channels;
see Figure \ref{diagram1.fig}) and iso-$\lambda$ nearly aligned with the detector X axis (mean tilt of 4$\pm 3^\circ$),
we make the additional
simplifying assumption to ignore this small tilt when computing the projected volume of the detector pixels
(see \S \ref{pipeline.sec}).  Effectively, this means that the surfaces
of the volume element are flat in the $\alpha$, $\beta$, and $\lambda$ planes\footnote{In making such an assumption, we are also implicitly ignoring the 
fact that the contents of the 3d volume represented by a given detector pixel are not uniformly scrambled.  
As for ordinary slit spectroscopy, spatial substructure within a given
slice in the across-slice ($\beta$) direction will be degenerate with
substructure in the spectral dispersion direction.}
and the spatial and spectral overlaps can be computed independently
(see Figure \ref{diagram2.fig}).

In the spectral domain, the overlap in $\lambda$ between detector
pixels and cube voxels is a simple 1d problem with a well-known
solution.
If the minimum and maximum wavelengths covered by the $i$th detector pixel are
given by $\lambda_{(i,1)}$ and $\lambda_{(i,2)}$ respectively, and the
minimum and maximum wavelengths covered by the $j$th cube voxel by $\lambda_{(j,1)}$ and $\lambda_{(j,2)}$, then the total length of overlap $\delta \lambda[i,j]$ in the spectral dimension between pixel $i$ and voxel $j$ is:

\begin{equation}
\begin{split}
\delta \lambda[i,j] = & \, \textrm{max}(0,\textrm{max}((\lambda_{(j,2)}-\lambda_{(i,1)}), \, 0) \\
& - \textrm{max}((\lambda_{(j,2)}-\lambda_{(i,2)}), \, 0)  \\
& - \textrm{max}((\lambda_{(j,1)}-\lambda_{(i,1)}), \, 0))
\end{split}
\end{equation}

In the spatial domain, we use the IFU spatial distortion transform
$(\ra,\dec) = \mathcal{D}(\alpha,\beta)$ mapping from local to world coordinates to
define the four corner coordinates $\mathcal{D}(\alpha_1,\beta_1)$, $\mathcal{D}(\alpha_2,\beta_1)$, $\mathcal{D}(\alpha_2,\beta_2)$, $\mathcal{D}(\alpha_1,\beta_2)$, where $\alpha_1$/$\alpha_2$ are the minimum/maximum along-slice coordinates associated with a given pixel and
$\beta_1$/$\beta_2$ are the minimum/maximum across-slice coordinates associated
with the corresponding slice.\footnote{Note that these coordinates pairs should wrap the footprint rather than crossing it on a 
diagonal since most standard area-overlap computations implicitly assume such
an ordering.}
Our exercise is thus to compute the common area $\delta \Omega[i,j]$ between the projected detector pixels defined by 
\begin{equation}
\begin{split}
    \Omega_i = &[(\ra_{(i,1)}, \dec{(i,1)}), (RA_{(i,2)}, DEC_{(i,2)}), \\
                & (RA_{(i,3)}, DEC_{(i,3)}), (RA_{(i,4)}, DEC_{(i,4)})]
\end{split}
\end{equation} 
and the cube spaxels defined by
\begin{equation}
\begin{split}
    \Omega_j = &[(RA_{(j,1)}, DEC_{(j,1)}), (RA_{(j,2)}, DEC_{(j,2)}), \\
                & (RA_{(j,3)}, DEC_{(j,3)}), (RA_{(j,4)}, DEC_{(j,4)})]
\end{split}
\end{equation} 
This problem can be solved using the well-studied Sutherland-Hodgman \citep{sh74} polygon
clipping algorithm.

Similarly to the imaging case,
we can likewise choose to shrink our effective pixel borders 
if desired by a fractional amount $p$
prior to computation of the common overlap area.  Doing so can improve the effective PSF FWHM of the drizzled
data cubes by about $5 - 10$\% 
\citep[i.e., in line with estimates of the PSF provided by 2d calibrated detector-level data; see discussion by][]{patapis23b},
but at the cost of making spectral resampling artifacts (\S \ref{resampling.sec}) substantially worse.

Adapting the formalism outlined by \citet{law16} for the SDSS-IV/MaNGA IFU,
we describe our input data as vectors of specific intensity
$f[i]$ (given for JWST in units of MJy sr$^{-1}$) and variance $g[i]$ for all
$i$ corresponding to valid detector pixels across all exposures to be
combined together.  Likewise, we define $F[j]$ and $G[j]$ to represent the 
specific intensity and variance of the $j$ voxels in the final 3d data cube.
We can thus write $F$ as the matrix product of $W$ and $f$

\begin{equation}
    F = W \times f
\end{equation}

where the normalized weights $W$ are given by
\begin{equation}
    W[i,j] = \frac{\delta \lambda[i,j] \, \delta \Omega[i,j]}{\sum_i \delta \lambda[i,j] \, \delta \Omega[i,j]}
\label{weights.eqn}
\end{equation}

The covariance matrix $C$ is given by
\begin{equation}
    C = W \times (g' \times W^\top)
\label{covar.eqn}
\end{equation}
where $g'$ is the variance matrix whose diagonal elements are the $g[i]$ and
all non-diagonal elements are zero.

Equivalently, the final voxel specific intensities and variances can also be
written in index notation as

\begin{equation}
    F[j] = \sum_i f[i] \, W[i,j]
\end{equation}

and 
\begin{equation}
    G[j] = \sum_i g[i] \, (W[i,j])^2
\end{equation}
respectively.

Once computed, these vectors $F$ and $G$ can be trivially rearranged 
into 3d arrays corresponding to the output data cube dimensions.
We note that this formalism is agnostic about the origin of the input data; while designed
for use with the JWST MIRI and NIRSpec IFUs, it can equally well be applied
to data obtained with ordinary slit-type instrument modes or even the NIRSpec microshutter array.  In this case the pixel
footprint would be defined by the pixel size in the along-slit direction
and the shutter width in the dispersion direction, and the same core algorithm
could be used to combine stepped-slitlet observations into composite data
cubes in future JWST observing cycles.

\subsection{Application to the JWST pipeline environment}
\label{pipeline.sec}



In terms of practical implementation in a pipeline environment, there is a
tension between the numerical accuracy of the solution and the CPU runtime
required to reach that solution.  With regard to our simplifying assumption that we
could separate the spatial and spectral components of the volume overlap
calculation, we compared two different 
pure python versions of the 3d drizzle algorithm; one in
which this simplification is made, and the other in which we calculate
the full volumetric overlap using a series of convex polyhedra.
The simplified code is found to run approximately 500 times faster, with a runtime
of about 15 minutes to build a data cube from a single MIRI MRS exposure
compared to approximately 120 hours using a standard Mac desktop.  The cubes themselves, however,
have statistically insignificant differences on the order of 0.03\% near pixels
whose trace tilt is approximately $10^{\circ}$ and approximately
0.003\% near pixels with a wavelength tilt of nearly $0^{\circ}$.
Given that the maximum MRS iso-$\lambda$ tilt is 12$^{\circ}$ and the maximum iso-$\alpha$ tilt is just
$3.3^{\circ}$, we conclude
that we can safely use the simplified method.\footnote{Likewise, the impact of this approximation on the reconstructed cubes will
be small compared to the instrumental systematics produced by fringing, PSF defocus, and the cruciform artifact \citep[see, e.g.,][and references therein]{argyriou23}.}

Even a 15-minute runtime for a simple single-frame data cube is impractical
however since cube building is run many times within the JWST pipeline architecture,
and typically with far more than a single exposure.
In contrast, we found that a python-wrapped C implementation of the core
3d drizzle algorithm reduced the runtime for a single-exposure MIRI MRS cube to just
7 seconds, representing a factor of $\sim$ 100 improvement with identical
result.


The optimal sampling parameters of the final data cube will vary as a function
of wavelength, and should be designed to approximately Nyquist sample both
the spatial point spread function (PSF) and the spectral line spread function
(LSF) in order to not lose information in the final data products.
Based on the observed PSF FWHM values for MIRI MRS data (see discussion
in \S \ref{ifuoverview.sec}), we adopt default values of $0.13\farcs$, $0.17\farcs$, $0.20\farcs$, and $0.35\farcs$ for the cube spaxel scales in Channels 1-4 respectively.  Likewise, based on the typical spectral resolving power $R$
within each band \citep{labiano21,jones23}, we adopt default spectral sampling steps for the four
channels of 
0.8~nm, 1.3~nm, 2.5~nm, and 6.0~nm 
respectively.
The JWST pipeline is thus able to construct data cubes with a common linear
spaxel scale and wavelength scale across all three grating settings within
each of the four MRS wavelength channels.

\begin{figure}[!]
\epsscale{1.2}
\plotone{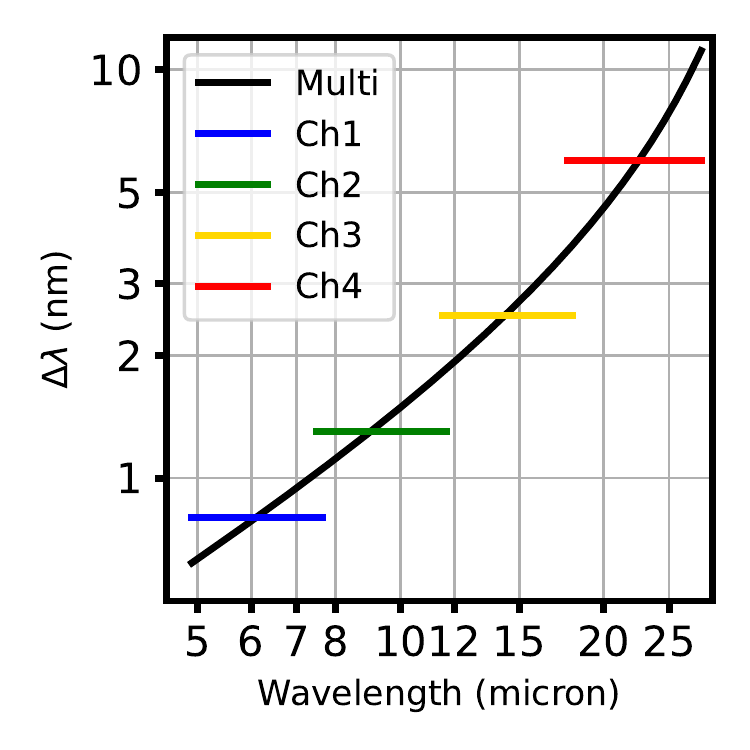}
\caption{MRS multi-channel cube wavelength sampling (solid black line), which varies as a function
of wavelength to approximately Nyquist sample the effective spectral resolution at all wavelengths.
Colored lines represent the constant linear sampling step used by default for cubes constructed
within a single MRS spectral channel.
}
\label{specres.fig}
\end{figure}

It is often convenient, however, to be able to construct composite drizzled data cubes across longer wavelength ranges combining multiple spectral channels in order to be able to effectively visualize
the corresponding spectra. Given the factor of $\sim 5$ in wavelength covered by MIRI MRS, however (and the factor of $\sim 8$ covered by NIRSpec)
this poses some challenges.  In particular, both the effective PSF FWHM
and spectral LSF change significantly over this range, meaning that the ideal cube voxel size
is different for each channel which is difficult to accommodate within the FITS
standard.  The pipeline therefore makes available a `multi-channel' drizzling option
in which the linear spatial sampling is set to the ideal
value for the shortest wavelengths and the non-linear spectral sampling is designed to 
provide roughly Nyquist
sampling of the spectral resolving power at all wavelengths (see Figure \ref{specres.fig}).
An example of a 1d spectrum extracted from such a multi-channel data cube was illustrated previously in Figure \ref{10lac.fig}
using a conical extraction radius that increases as a function of wavelength to account for the increasing PSF FWHM.

While the 3d drizzle algorithm can thus combine data from multiple
spectral bands (and even, in principle, between MIRI MRS and NIRSpec) we
note that this can complicate the analysis of the data within the spectral overlap regions between different bands.
In these regions cube building may combine together data with quite different spatial and/or spectral resolutions.  In the spatial domain, the different slicer widths between channels can produce a discontinuous jump in the central PSF structure and require correspondingly different aperture correction factors for small extraction radii.  In the spectral domain such a combination of data from different bands can produce a hybrid LSF, although as 
discussed by \citet[][see their Figure 7]{law21}
for relatively small LSF differences ($\sim$ 20\%) such a combination can be described to reasonable accuracy as the simple mean of the
two LSF widths.



\section{Covariance}
\label{covar.sec}

Since the process of building a rectified 3d data cube involves resampling the original data, it necessarily
introduces covariance into the resulting cubes.  This covariance between individual voxels means
that the variance of spectra extracted from these data cubes does not improve with increasing aperture radius as quickly as 
might otherwise be expected, and is thus important to take into consideration in the analysis of IFU spectra.

Formally, the covariance of JWST data cubes can be computed for an arbitrary set of observations using Equation~\ref{covar.eqn}.
For a typical MRS Channel 1 data cube (dimensions $\sim 50 \times 50 \times 3500$), the corresponding covariance matrix has
nearly $10^{14}$ elements describing the correlation between every pair of voxels, requiring roughly 0.3 PB to store.
Fortunately this matrix is extremely sparse for the 3d drizzle weighting function, with all non-zero entries closely confined to within
a few elements of the diagonal.  In the future, the JWST pipeline may provide such sparse matrices
since they can be useful in computing SNR statistics for binned regions of the data cubes \citep[see, e.g., discussion 
for the MaNGA IFU data products in Section 6 of][]{westfall19}.

In the present contribution, we simply estimate an approximate scaling factor that can be applied to the variance computed
for 1d extracted spectra as a function of the aperture extraction radius for typical MIRI MRS data cubes.
Such an approach has a long history of use in the literature, and was adopted for both the CALIFA IFU survey \citep{husemann13} and early
MaNGA IFU survey \citep{law16}, in addition to classical HST imaging applications \citep{casertano00,fh02}.

We determine this scaling factor empirically by taking a set of 4-pt dithered MRS observations,
replacing all ERR extension values with an arbitrary constant value of 0.1 MJy/sr, and replacing all 
SCI extension values with quantities
drawn from a gaussian random distribution centered around 1 MJy/sr with a $1\sigma$ distribution width of 0.1 MJy/sr.  We then construct
science and error cubes from the data following the prescriptions in \S \ref{drizzle.sec} 
and the default cube parameters given in \S \ref{pipeline.sec}
to obtain a single realization of the composite
cube.  We repeat this exercise 20 times with a different seed for the random number generator.
For every voxel in the data cube, we can thus directly compare the pipeline-estimated variance in the voxel with the actual variance observed
between the 20 random trials.  Similarly, for arbitrary aperture sizes we can extract 1d spectra from the data cubes
and compare the nominal diagonal variance in the spectrum $\sigma_{\rm diag}$ to the effective variance $\sigma_{\rm meas}$ observed between spectra extracted from the 20 random realizations.

\begin{figure}[!]
\epsscale{1.2}
\plotone{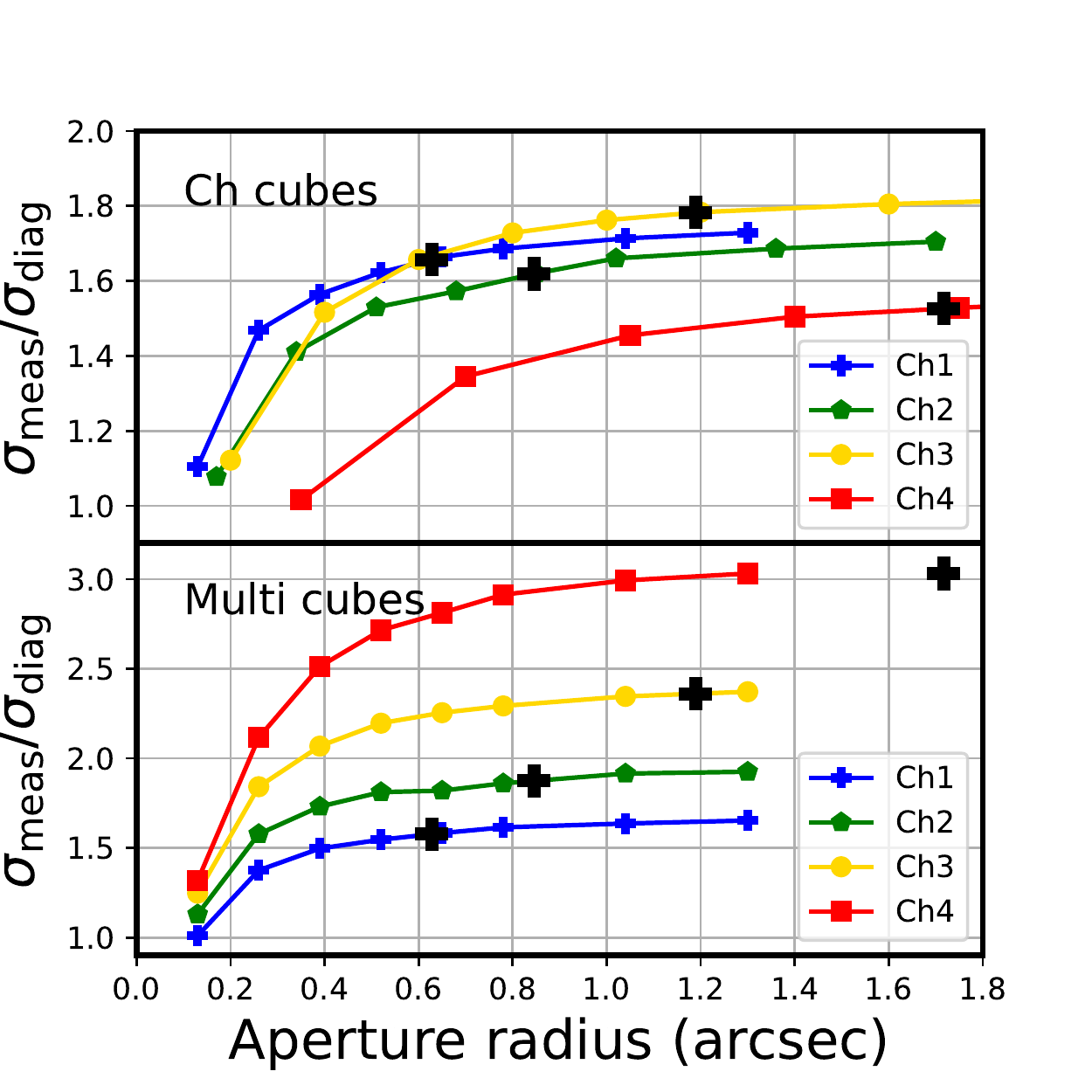}
\caption{Ratio of the error measured from multiple simulated realizations of 
constant surface brightness data cubes ($\sigma_{\rm meas}$) divided by the error estimated
from the cube error array assuming a diagonal covariance matrix ($\sigma_{\rm diag}$; i.e., no
covariance between adjacent spaxels).  The top panel shows the results as a function of the 
aperture extraction radius for Channels 1-4 when each channel has a voxel scale as given in 
\S \ref{pipeline.sec}.  The bottom panel shows similar results when all four
channels are built onto a single output cube scale with spatial sampling optimized for Ch1
and a non-linear wavelength solution.  Solid black plus signs represent the interpolated values
at radii 2.0 times the PSF FWHM.
}
\label{covar.fig}
\end{figure}

We show the results of this experiment in Figure \ref{covar.fig} for both channel-specific IFU cubes and multi-channel IFU cubes
spanning the entire MRS wavelength range.\footnote{There is no significant dependence on location within the FOV or wavelength within  a given channel.}
Figure \ref{covar.fig} demonstrates that the ratio $\sigma_{\rm meas}/\sigma_{\rm diag}$ increases asymptotically with increasing aperture radius.  For channel-specific
cubes, this ratio is 1.66, 1.62, 1.78, and 1.53 at $r = 2.0$ FWHM for Channels 1-4 respectively (default spaxel scales 0.13, 0.17, 0.20, and 0.35 arcsec).  The corresponding ratios for the multi-channel cubes (1.58, 1.87, 2.36, and 3.03 for Channels 1-4 respectively) increase significantly at longer wavelengths due to the significant oversampling of the larger PSF by a spaxel scale designed for short wavelengths.
We therefore recommend that scientific analyses using spectra extracted from the JWST
data cubes scale their variance accordingly in order to account for this effect.


\section{Consequences of Undersampling}
\label{resampling.sec}

As discussed in \S \ref{ifuoverview.sec}, the MRS is significantly undersampled
compared to the ideal Nyquist frequency.  As a result, resampling the 2d detector data
into 3d rectified data cubes produces aliasing artifacts 
\citep[aka resampling noise, see][]{smith07}
in the final data products.\footnote{This resampling noise is in addition to the large ($\sim 30$\%) periodic amplitude modulations produced in
MIRI spectra due to fringing within the instrument \citep[see discussion by][and references therein]{argyriou23}.}
While dithering helps to mitigate these artifacts by filling in the missing phase space
it does not eliminate them entirely.

\begin{figure*}[!]
\epsscale{1.1}
\plotone{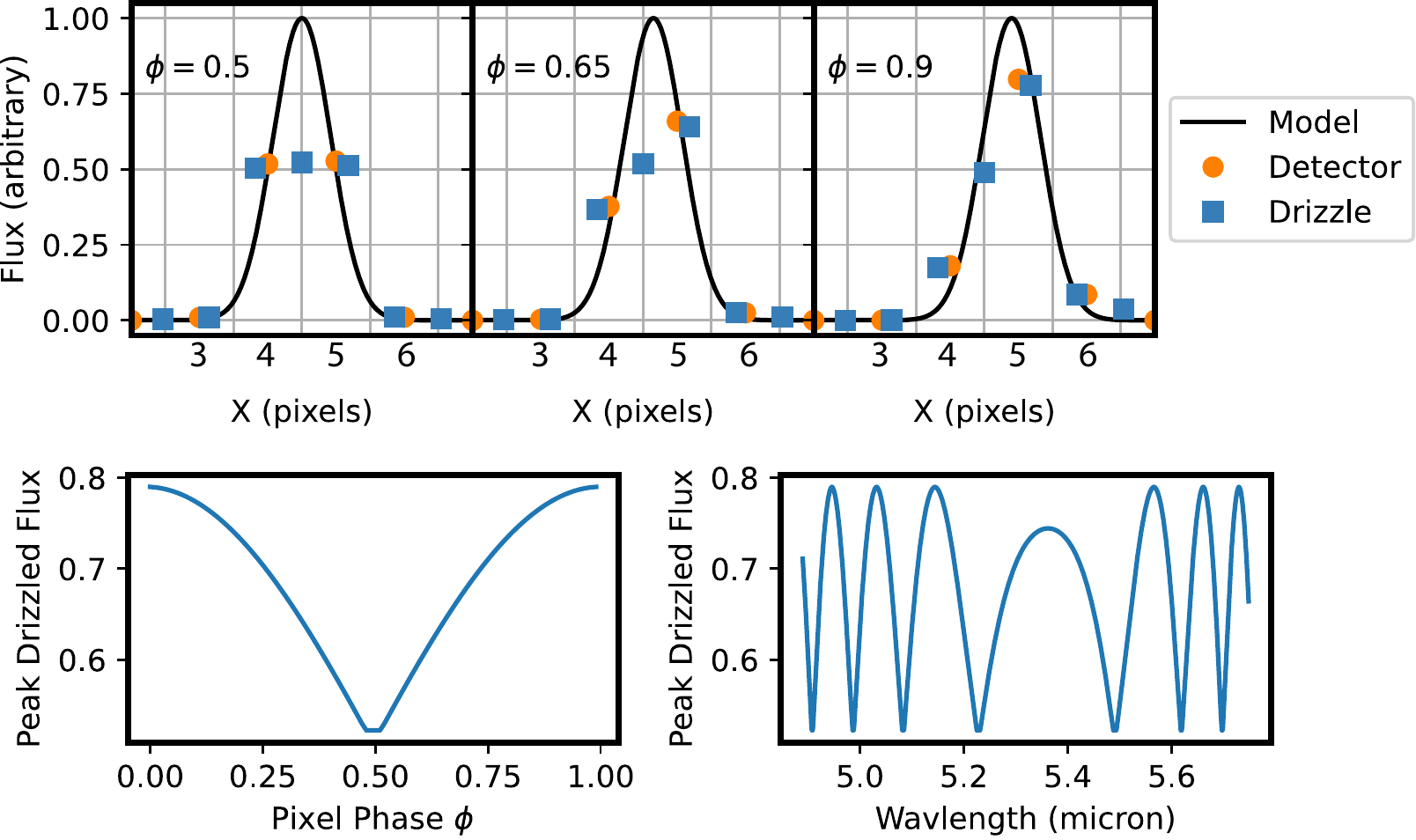}
\caption{One-dimensional illustration of the origins of resampling noise.
The top panel shows three realizations of a 1d gaussian line profile with different peak locations.  Orange circles represent the values recorded after convolution of the line profile with the tophat pixel boundaries (indicated by the vertical grid lines), and the blue square the values after drizzling to a different sampling scale.  In each case $\phi$ gives the pixel phase of the gaussian peak.
Bottom left: Peak drizzled flux value as a function of the pixel phase of the gaussian peak.  Bottom right: Peak drizzled flux as a function of wavelength assuming a pixel phase that varies with wavelength in a manner similar to a typical MRS spectral trace.
}
\label{resamp1d.fig}
\end{figure*}

We illustrate the origin of these artifacts schematically using a 1d
example in Figure \ref{resamp1d.fig}.
In this figure we plot a 1d gaussian function representing an intrinsic
signal (black line in upper panels) and convolve that function with a tophat pixel response function to determine the counts that would be recorded in the pixelized representation
of that function (filled orange circles).  The pixel width is here chosen to be roughly half-Nyquist,
similar to the case of the MRS at many wavelengths.  The pixelized representation of the
function is then resampled to a different pixel grid using a 1d version
of the drizzle algorithm (filled blue squares).  While the integrated counts in the resampled
function are independent of the phase $\phi$ of the intrinsic signal peak with respect
to the pixel boundaries, the profiles change dramatically.  The flux in the peak
resampled pixel for instance varies from 50\% to 80\% of the intrinsic peak flux; as
illustrated in the lower-left panel this value varies systematically with the pixel phase.
If we assume that the pixel phase of the signal peak varies with wavelength in a manner
akin to the trace of the spectral dispersion on the MRS detectors in Channel 1A, we would
therefore expect a peak resampled count rate that shows a sinusoidal oscillation with
wavelength (lower-right panel).  The effective frequency of this sinusoidal oscillation
changes with wavelength according to the tilt between the spectral dispersion and the detector
pixel grid; at the ends of the detector where the trace crosses pixel boundaries faster
the oscillation frequency is higher than near the middle of the detector where the trace
is nearly aligned with the detector Y axis.

\begin{figure*}[!]
\epsscale{1.1}
\plotone{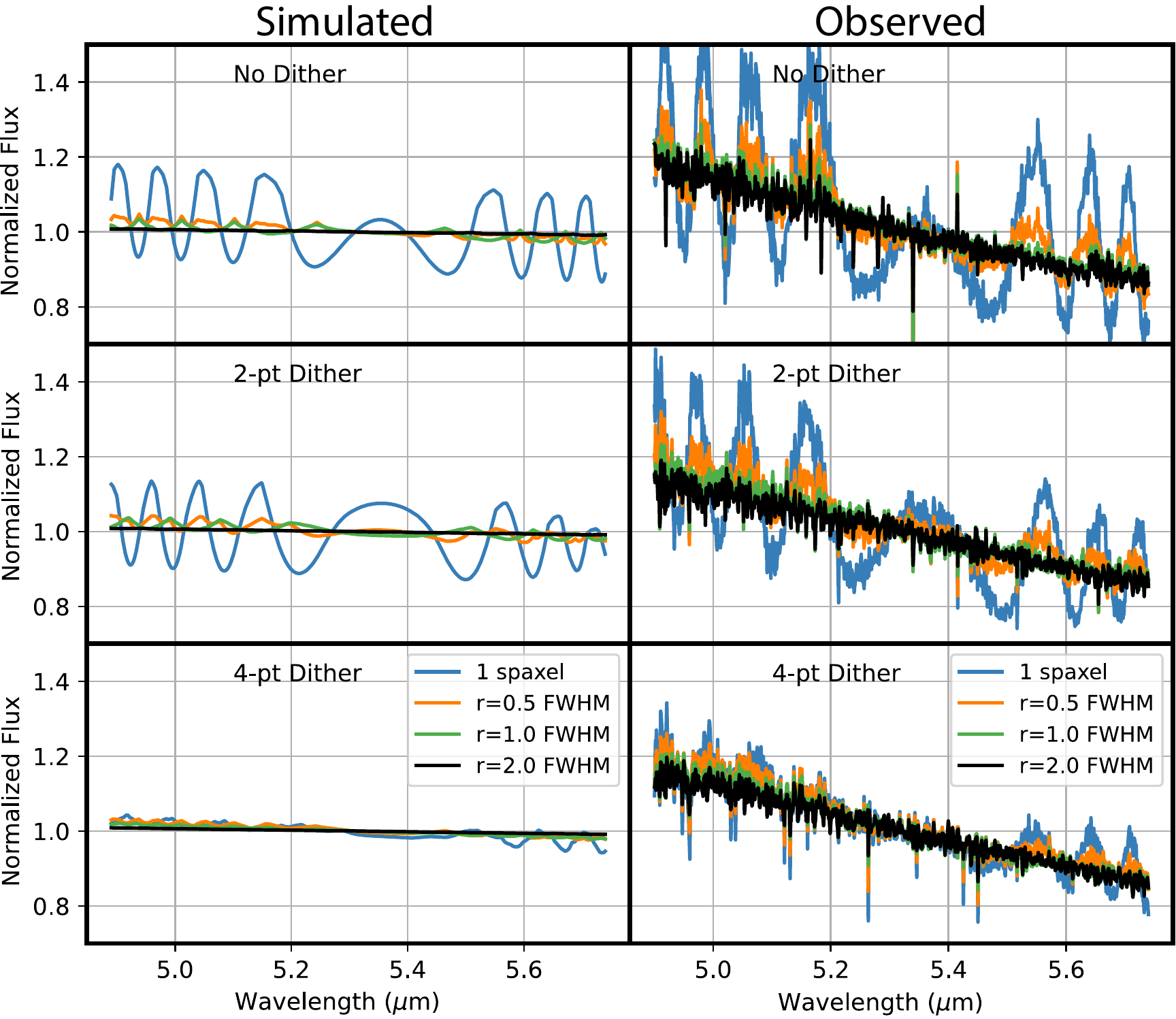}
\caption{Simulated (left) and observed (right) effect of resampling noise on extracted MRS spectra from drizzled data cubes.
Simulations assumed a flat noiseless spectrum, while observations targeted the bright G3V star 16CygB.  Resampling noise imparts large
low-frequency modulations in the spectrum, and is mitigated by dithering and use of large extraction apertures.
}
\label{resamp.fig}
\end{figure*}

In three dimensions the situation is slightly more complex but the effect broadly equivalent.
In Figure \ref{resamp.fig} (left-hand panels) we show the results of numerical simulations
in which we constructed a detailed model of the JWST PSF using the WebbPSF tool \citep{perrin14}, and
simulated a source with constant 1 Jy spectrum with a realistic model of the MRS distortion
solution and flight-like dither patterns.  These simulations neglected all sources of noise and detector responsivity variations (e.g., fringing, spectrophotometric calibration, etc) and considered
only the effective sampling of each detector pixel based on the instrument distortion model.
These dithered data were combined into a 3d data cube using the drizzle algorithm defined
in \S \ref{drizzle.sec}.
The spectrum of a single cube spaxel (solid blue line) shows the same kind of wavelength-dependent amplitude variations as seen in the simplified 1d model shown in Figure \ref{resamp1d.fig}.

These expectations are confirmed by in-flight observations.  In Figure \ref{resamp.fig} (right-hand panels) we plot spectra extracted from Cycle 1 calibration observations of the 6th-magnitude G3V star 16CygB (JWST Program ID 1538).  These spectra show similar large-scale amplitude
variations with comparable changes in frequency as a function of wavelength.

Such behaviour has been observed before in data from previous generations of space-based telescopes.
\citet{anderson16} for instance discussed the impact of such pixel-phase effects on the
reconstruction of the point spread function for HST/WFC3 imaging data, while \citet{dressel07}
discussed the effect on HST/STIS spectra whose spectral traces change pixel phase rapidly.
The JWST/MIRI IFU case is akin to a combination of the two; plotting the spectrum of a single
cube spaxel is similar to trying to compare source fluxes in imaging data by plotting the peak
value of the drizzled image for each source rather than doing proper aperture photometry.\footnote{Such effects are less pronounced in data cube produced by fiber-fed spectrographs such as MaNGA because of azimuthal scrambling in combination with the relevant
sampling taking place at the fiber face whose footprint changes only slowly with wavelength
due to chromatic differential refraction.}

As for the imaging and single slit cases, the solution to this problem for IFU spectroscopy
is a combination of dithering and a suitably large spectral extraction aperture.
As illustrated by Figure \ref{resamp.fig}, the amplitude of this 'resampling noise' decreases
with the use of a four-point dither pattern that samples the PSF at roughly
half-integer intervals.  A two-point pattern in contrast does not significantly improve
resampling noise beyond the undithered case, largely because optical distortions
in the MRS combined with pixel-mapping discontinuities between slices
prohibits such a simple pattern from actually achieving half-integer sampling
throughout the entire field of view.
Similarly, resampling noise decreases dramatically with the use of larger extraction
apertures.  Even undithered data for instance show little to no resampling noise when
using apertures whose radius is 2.0 times the PSF FWHM.  This is unsurprising, given
we know that the total {\it integrated} flux is conserved.

\begin{figure}[!]
\epsscale{1.2}
\plotone{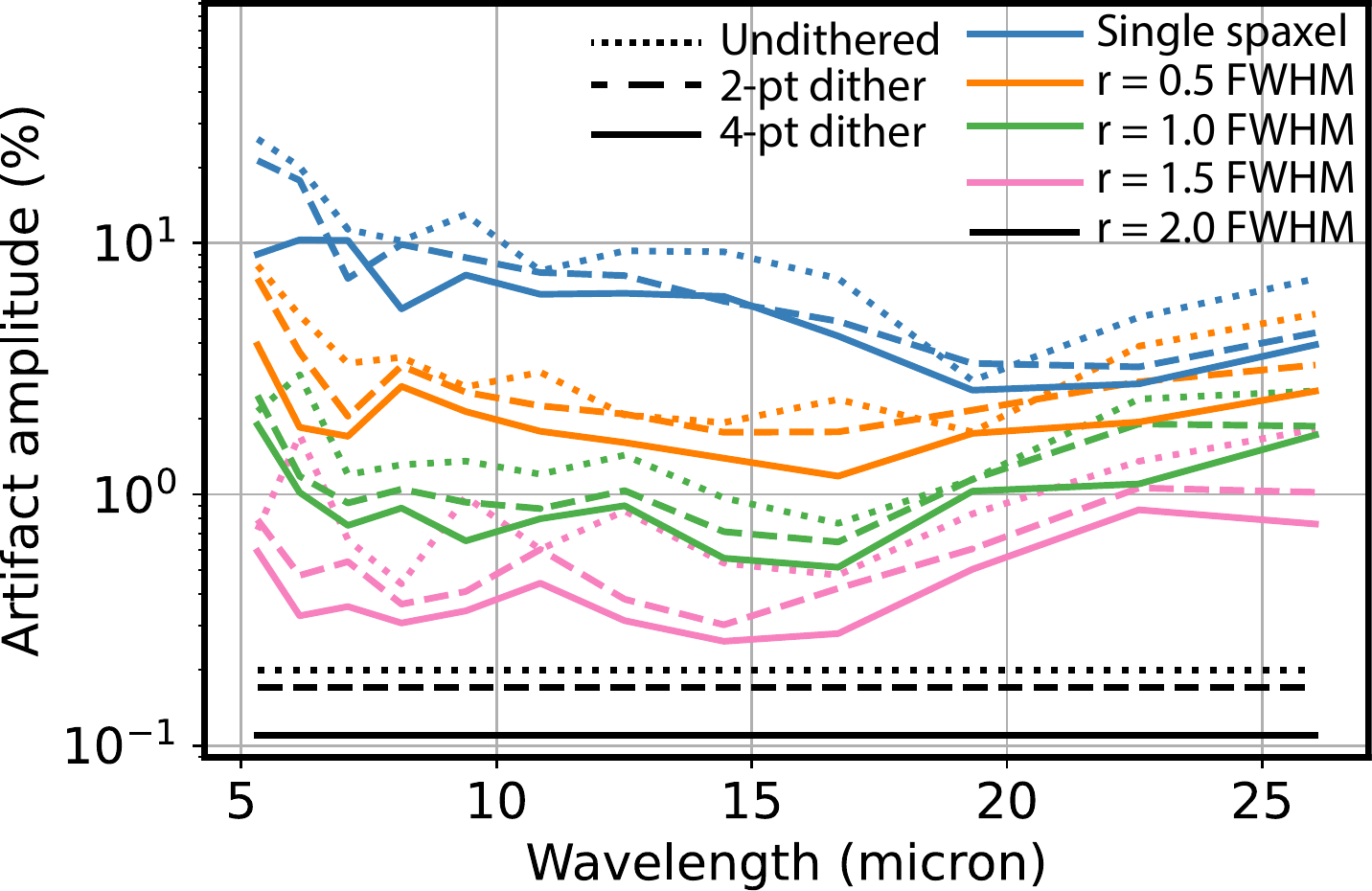}
\caption{Semi-amplitude of resampling artifacts (see Figure \ref{resamp.fig}) in 1d spectra extracted from observations of
the bright star 16CygB as a function of wavelength.
Dotted/dashed/solid lines respectively indicate undithered data, data obtained with a 2-pt dither, and data obtained with a 4-pt dither respectively.  Spectra are extracted from single spaxels (cyan lines), apertures with radius 0.5 times the PSF FWHM (orange lines), 1.0 times the PSF FWHM (green lines), 1.5 times the PSF FWHM (pink lines), and 2.0 times
the PSF FWHM (black lines).  Both dithering and larger aperture radii help reduce the amplitude of resampling artifacts in extracted spectra; the apparent rise in artifact amplitudes at the longest wavelengths is due in part to the increasing noise in the observed spectrum.}
\label{resamp_percent.fig}
\end{figure}

We estimate the impact of this resampling noise on spectra throughout the MRS wavelength
range using MIRI Cycle 1 calibration observations of bright G3V star 16CygB
(JWST Program ID 1538).
In each of the twelve MRS bands, we extract spectra from the composite data cubes using apertures of radius 0\footnote{I.e., a single spaxel.}, 0.5,  1.0, 1.5, and 2.0 times the PSF FWHM given by Eqn \ref{psf.eqn}.
We apply a 1d residual fringe correction \citep{kavanagh23} to our extracted spectra in order to remove periodic modulations of known frequency produced by fringing within the MIRI instrument that are unrelated to any sampling considerations \citep[see discussion by][and references therein]{argyriou23}.  We then compute the normalized ratio between spectra extracted from the smaller apertures to
spectra extracted from the $r = 2$ FWHM aperture (the default for the MIRI MRS pipeline), and measure 
the semi-amplitude of the resampling-induced oscillation.\footnote{We do not apply the aperture correction factors necessary to obtain the true flux from each aperture, as our normalization would eliminate such global correction factors anyway.}
These values are plotted in 
Figure \ref{resamp_percent.fig} for cubes reconstructed from single undithered observations, two-point
dithered data, and four-point dithered data.
In order to reduce sampling artifacts in the recovered spectra to 5\% or less at all wavelengths we recommend use of a 4-point dither pattern with an extraction radius
of at least 0.5 times the PSF FWHM, while reducing such artifacts
below 1\% requires an extraction radius of at least 1.5 times the PSF FWHM.\footnote{As a practical consequence, resampling noise will be larger in the preview cubes produced by the {\sc calwebb\_spec2} stage of the JWST pipeline which incorporate just a single exposure than in the {\sc calwebb\_spec3} data cubes combining multiple dithered exposures.}

\section{Conclusions}
\label{summary.sec}

We have presented an algorithm for the extension of the classic 2d
drizzle technique to three dimensions for application to the MIRI and NIRSpec integral
field unit spectrometers aboard JWST.  While this technique is conceptually straight forward and relies
on a weight function proportional to the volumetric overlap between native detector pixels
and resampled data cube voxels, in practice additional simplifications are necessary to
make the problem computationally tractable.  By separating the spatial and spectral
overlap computations it is possible to achieve orders of magnitude gains in computational time at
minimal cost in spectrophotometric accuracy.  Such gains, along with speed improvements
provided by implementation of the core algorithm in a pre-compiled (i.e, C-based)
architecture are essential for implementation of this approach in a practical
pipeline environment.

While this algorithm provides a matrix-based formalism for computation of the variance
of the final cube spaxels and the covariance between such spaxels, the full covariance
matrix is intractably large for practical use.  This covariance means that
the effective variance in extracted spectra does not average down with the inclusion
of more cube spaxels in the naively expected manner.  We therefore computed a series of
simple scaling relations that can be applied to spectra extracted from MIRI MRS
data cubes in order to account for covariance between cube spaxels.  Depending on the 
wavelength band in question, the cube sampling, and the aperture radius, such correction factors range from $\sim 1.5$ to 3.

Finally, we discussed the practical implications of the severe (factor $\sim 2$) 
undersampling of the MIRI MRS on the quality of the calibrated data cubes provided by the
3d drizzle technique.  We demonstrated that this undersampling imparts a periodic 
amplitude modulation of up to 20\% into spectra extracted from such data cubes without
efforts to mitigate it.  With the use of appropriate dithering (at least 4 points)
and spectral aperture extraction radii ($r = 1.5$ FWHM or greater) these sampling artifacts
can be reduced to 1\% or less throughout the MRS spectral range.


\begin{acknowledgments}

We thank Eric Emsellem and Peter Weilbacher for helpful discussions with regard to IFU cube building applications to the WHT/SAURON and VLT/MUSE integral field spectrographs.

Ioannis Argyriou and Bart Vandenbussche thank the European Space Agency (ESA) and the Belgian Federal Science Policy Office (BELSPO) for their support in the framework of the PRODEX Programme.

J.A.-M. acknowledge support by grant PIB2021-127718NB-100 from the Spanish Ministry of Science and Innovation/State Agency of Research MCIN/AEI/10.13039/501100011033 and by “ERDF A way of making Europe”.

The following National and International Funding Agencies funded and supported the MIRI development: NASA; ESA; Belgian Science Policy Office (BELSPO); Centre Nationale d'Etudes Spatiales (CNES); Danish National Space Centre; Deutsches Zentrum fur Luftund Raumfahrt (DLR); Enterprise Ireland; Ministerio De Economi´a y Competividad; Netherlands Research School for Astronomy (NOVA); Netherlands Organisation for Scientific Research (NWO); Science and Technology Facilities Council; Swiss Space Office; Swedish National Space Agency; and UK Space Agency.

This manuscript uses data from DOI 10.17909/63a3-d535

\end{acknowledgments}

\appendix

\section{Application to JWST/NIRSpec}
\label{nrs_appendix.sec}

While our analysis in this contribution has focused primarily on the JWST/MIRI MRS IFU, all of the considerations of resampling noise and covariance likewise apply to data obtained with the JWST/NIRSpec IFU as well.  We do not explore all operational modes of NIRSpec in detail, but present here a brief analysis of a single band to demonstrate this similarity.  For this comparison we use observations of the standard star GSPC P330-E (spectral type G2V) obtained by JWST Program ID 1538 (Observation 62) in the G140H/F100LP grating using a 4-point nod dither pattern.

\begin{figure}[!]
\epsscale{1.2}
\plotone{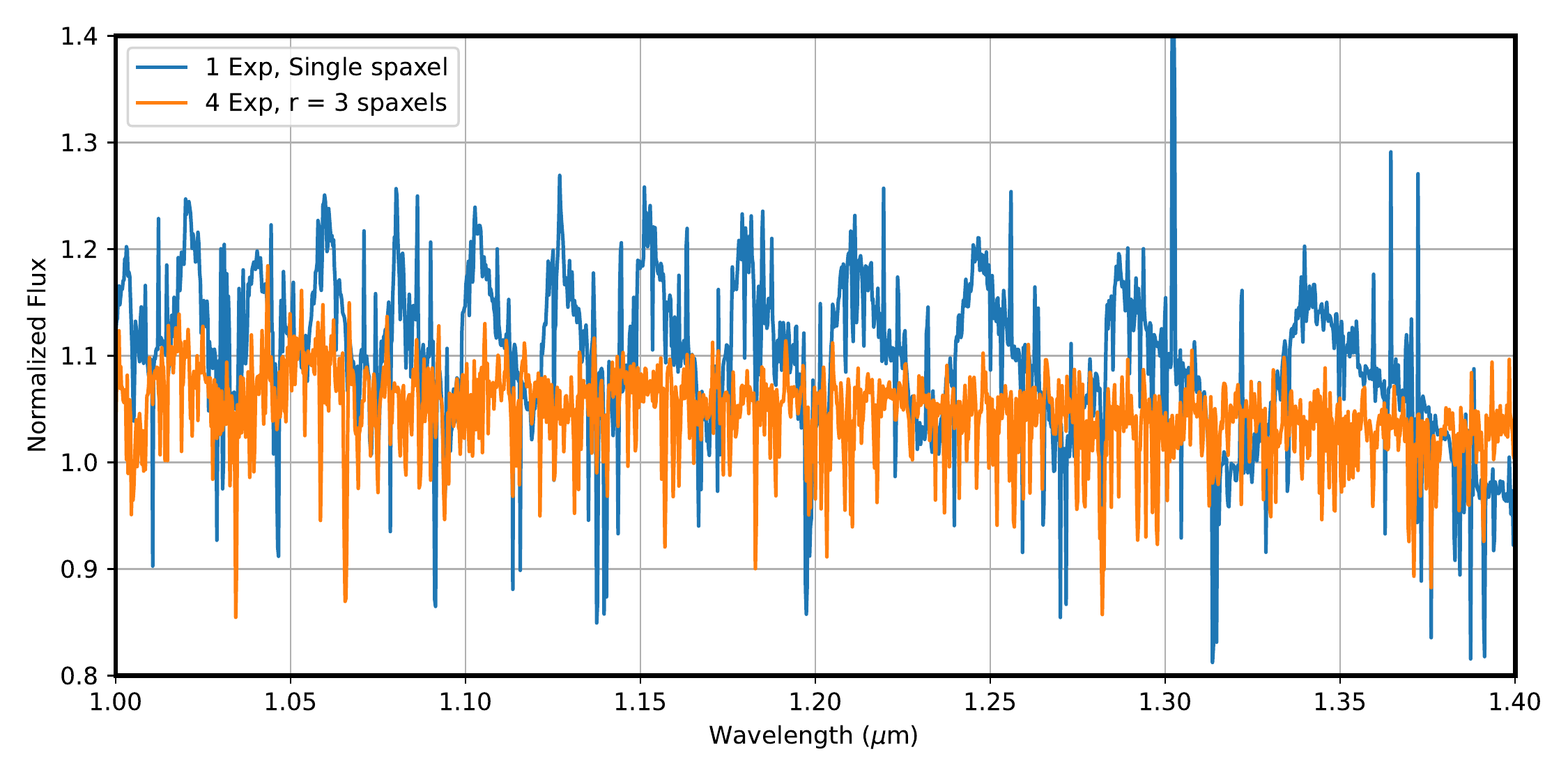}
\caption{JWST/NIRSpec G140H/F100LP spectrum of G2V star GSPC P330-E.  The solid blue line shows the spectrum of the brightest spaxel in a single exposure, and exhibits significant resampling noise compared to the orange spectrum extracted from a 3-spaxel radius region from a cube built from four dithered exposures.}
\label{nirspec.fig}
\end{figure}

After processing the observations through the JWST pipeline we use the 3D drizzle algorithm to build data cubes from individual exposures and from the composite 4-point dither sequence.  In Figure \ref{nirspec.fig} we plot the resulting spectrum in the wavelength range $\lambda\lambda 1.00-1.40 \mu$m for a single spaxel located in the center of the star for the individual-exposure cube (solid blue line) and from a three-spaxel radius aperture around the star in the 4-point dithered data cube (solid orange line).
As for the MIRI MRS observations, we note a strong beating due to resampling noise in the single-spaxel spectrum whose frequency varies with wavelength according to the rate at which the dispersed spectral traces cross detector pixel boundaries.  This beating is significantly reduced in the spectrum extracted from dithered data with a moderate aperture extraction radius.

Likewise, if we repeat our analysis from \S \ref{covar.sec} with the NIRSpec data we can estimate the covariance between spaxels in the rectified data cubes.  In the case of G140H/F100LP we find a covariance factor of about 1.24 for apertures three spaxels in radius.  This factor is somewhat lower than shown for typical MIRI MRS observations in Figure \ref{covar.fig} because the spaxel scale of the
data cube (0.1 arcsec by default) is comparable to the native pixel scale and slice width with which the scene was originally sampled (in contrast, the MIRI MRS Channel 1 default spaxel scale is about 70\% of the native sampling size in order to maximize the angular resolution of the reconstructed cubes).
If the NIRSpec cubes were reprocessed with a finer spaxel scale the covariance would increase accordingly.

\section{Exponential Modified Shepard Method (EMSM)}
\label{emsm.sec}

As noted in \S \ref{ifuoverview.sec}, the JWST pipeline also contains a second independent method of building data cubes.  This second method uses a flux-conserving variant of Shepard's interpolation method commonly employed for ground-based IFUs \citep[e.g., for the SDSS-IV/MaNGA survey, see discussion by][]{law16}.  In brief, each detector pixel is described by its midpoint position in (\ra, \dec, $\lambda$) and combined into the final data cube using a 3d weighting kernel $w$ described by

\begin{equation}
    w = e^{-(x^2 + y^2 + z^2)/\sigma^2}
\end{equation}
where $x, y, z$ are normalized distances in voxel units in the three axes of the data cube (two spatial axes and one spectral) and $\sigma$ is the scale radius of the exponential profile.  Since this function formally extends to infinity, it must be truncated within some limiting region of influence in both spatial and spectral dimensions and normalized by the sum of individual weights to ensure flux conservation.

While the 3d drizzle approach has no free parameters (other than the desired output voxel size), EMSM thus has many such parameters including the scaling and truncation radii in all three dimensions that must be tuned to provide reasonable performance for each
spectral band.  Effectively, while the 3d drizzle method apportions flux from a given pixel evenly to the region defined by the boundaries of that pixel, the EMSM method apportions flux to an exponential profile whose size and shape can be tuned by the user.

\begin{figure}[!]
\epsscale{1.2}
\plotone{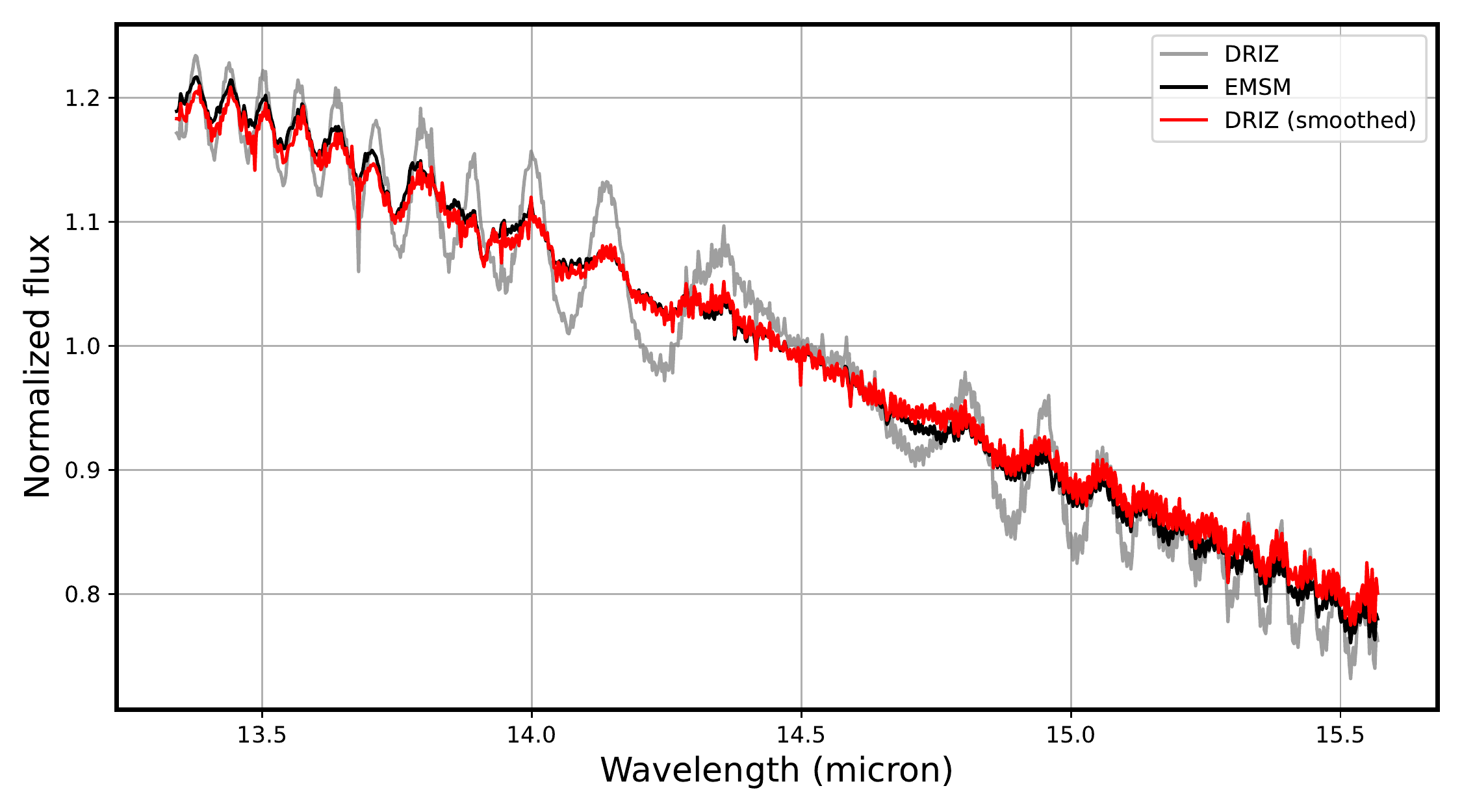}
\caption{MIRI MRS Channel 3B spectrum of standard star 16 CygB, extracted from the single brightest spaxel in a data cube constructed from four dithered exposures.  The solid grey and black lines show the result from data cubes constructed using the 3d drizzle and EMSM methods respectively, while the red line shows the result from the drizzled data cube smoothed with a spatial gaussian kernel of radius 0.7 spaxels prior to spectral extraction.}
\label{emsm.fig}
\end{figure}

In practice, data cubes constructed using the EMSM approach have marginally lower spatial and spectral resolution than their drizzled counterparts due to the introduction of the exponential convolution kernel which typically extends beyond the native pixel boundaries.  This smoothing can also result in differences in the effective resampling noise between cubes constructed using the drizzle vs EMSM methods.
In Figure \ref{emsm.fig} we illustrate one such example from the MRS Channel 3B observations of 16CygB; in this case the brightest-spaxel spectrum obtained from the EMSM data cube (black line) shows resampling artifacts that are about half the amplitude of those in the drizzled data cube (grey line).  Similar results can be obtained from the drizzled data cube by convolving it with a spatial gaussian kernel comparable to the EMSM scale factor $\sigma$ (Figure \ref{emsm.fig}, red line) prior to spectral extraction.

\begin{thebibliography}


\bibitem[Anderson(2016)]{anderson16} Anderson, J.\ 2016, Instrument Science Report WFC3 2016-12, 42 pages

\bibitem[Argyriou et al.(2020)]{argyriou20} Argyriou, I., Rieke, G.~H., Ressler, M.~E., et al.\ 2020, \procspie, 11454, 114541P. doi:10.1117/12.2561502

\bibitem[Argyriou et al.(2023)]{argyriou23} Argyriou, I., Glasse, A., Law, D.~R., et al.\ 2023, arXiv:2303.13469. doi:10.48550/arXiv.2303.13469

\bibitem[B{\"o}ker et al.(2022)]{boker22} B{\"o}ker, T., Arribas, S., L{\"u}tzgendorf, N., et al.\ 2022, \aap, 661, A82. doi:10.1051/0004-6361/202142589


\bibitem[Bushouse et al.(2022)]{bushouse22} Bushouse, H., Eisenhamer, J., Dencheva, N., et al.\ 2022, Zenodo


\bibitem[Casertano et al.(2000)]{casertano00} Casertano, S., de Mello, D., Dickinson, M., et al.\ 2000, \aj, 120, 2747. doi:10.1086/316851

\bibitem[Dressel et al.(2007)]{dressel07} Dressel, L., Hodge, P., \& Barrett, P.\ 2007, Instrument Science Report STIS 2007-04, 20 pages


\bibitem[Fruchter \& Hook(2002)]{fh02} Fruchter, A.~S. \& Hook, R.~N.\ 2002, \pasp, 114, 144. doi:10.1086/338393




\bibitem[Husemann et al.(2013)]{husemann13} Husemann, B., Jahnke, K., S{\'a}nchez, S.~F., et al.\ 2013, \aap, 549, A87. doi:10.1051/0004-6361/201220582

\bibitem[Jones et al.(2023)]{jones23} Jones, O.~C., {\'A}lvarez-M{\'a}rquez, J., Sloan, G.~C., et al.\ 2023, arXiv:2301.13233. doi:10.48550/arXiv.2301.13233

\bibitem[Kavanagh et al. (2023)]{kavanagh23} Kavanagh, P.\ et al. 2023, in prep.

\bibitem[Labiano et al.(2021)]{labiano21} Labiano, A., Argyriou, I., {\'A}lvarez-M{\'a}rquez, J., et al.\ 2021, \aap, 656, A57. doi:10.1051/0004-6361/202140614

\bibitem[Law et al.(2016)]{law16} Law, D.~R., Cherinka, B., Yan, R., et al.\ 2016, \aj, 152, 83 

\bibitem[Law et al.(2021)]{law21} Law, D.~R., Westfall, K.~B., Bershady, M.~A., et al.\ 2021, \aj, 161, 52. doi:10.3847/1538-3881/abcaa2


\bibitem[Morrison et al. (2023)]{morrison23} Morrison, J.\ et al. 2023, in prep.

\bibitem[Mueller et al. (2023)]{mueller23} Mueller, M.\ et al. 2023, in prep.

\bibitem[Patapis et al. (2023)]{patapis23a} Patapis, P.\ et al. 2023a, in prep.

\bibitem[Patapis et al. (2023)]{patapis23b} Patapis, P.\ et al. 2023b, in prep.

\bibitem[Perrin et al.(2014)]{perrin14} Perrin, M.~D., Sivaramakrishnan, A., Lajoie, C.-P., et al.\ 2014, \procspie, 9143, 91433X. doi:10.1117/12.2056689

\bibitem[Regibo(2012)]{regibo12} Regibo, S.\ 2012, Ph.D. Thesis

\bibitem[Ressler et al.(2015)]{ressler15} Ressler, M.~E., Sukhatme, K.~G., Franklin, B.~R., et al.\ 2015, \pasp, 127, 675. doi:10.1086/682258

\bibitem[Rieke et al.(2015)]{rieke15} Rieke, G.~H., Ressler, M.~E., Morrison, J.~E., et al.\ 2015, \pasp, 127, 665. doi:10.1086/682257

\bibitem[S{\'a}nchez et al.(2012)]{sanchez12} S{\'a}nchez, S.~F., Kennicutt, R.~C., Gil de Paz, A., et al.\ 2012, \aap, 538, A8. doi:10.1051/0004-6361/201117353


\bibitem[Sharp et al.(2015)]{sharp15} Sharp, R., Allen, J.~T., Fogarty, L.~M.~R., et al.\ 2015, \mnras, 446, 1551. doi:10.1093/mnras/stu2055

\bibitem[Smith et al.(2007)]{smith07} Smith, J.~D.~T., Armus, L., Dale, D.~A., et al.\ 2007, \pasp, 119, 1133. doi:10.1086/522634

\bibitem[Sutherland \& Hodgman(1974)]{sh74} Sutherland, I.~E. \& Hodgman, G.~W.\ 1974, ``Reentrant polygon clipping'', Commun. ACM, Vol. 17, No. 1, pp. 32-42.

\bibitem[Weilbacher et al.(2020)]{weilbacher20} Weilbacher, P.~M., Palsa, R., Streicher, O., et al.\ 2020, \aap, 641, A28. doi:10.1051/0004-6361/202037855

\bibitem[Wells et al.(2015)]{wells15} Wells, M., Pel, J.-W., Glasse, A., et al.\ 2015, \pasp, 127, 646. doi:10.1086/682281




\bibitem[Westfall et al.(2019)]{westfall19} Westfall, K.~B., Cappellari, M., Bershady, M.~A., et al.\ 2019, \aj, 158, 231. doi:10.3847/1538-3881/ab44a2

\bibitem[Wright et al. (2023)]{wright23} Wright, G.\ et al. 2023, PASP submitted.



\end{thebibliography}
\end{document}